\PassOptionsToPackage{table}{xcolor}
\documentclass[sigconf]{acmart}

\AtBeginDocument{%
  }

\copyrightyear{2025}
\acmYear{2025}
\setcopyright{cc}
\setcctype{by}
\acmConference[CCS '25]{Proceedings of the 2025 ACM SIGSAC Conference on Computer and Communications Security}{October 13--17, 2025}{Taipei, Taiwan}
\acmBooktitle{Proceedings of the 2025 ACM SIGSAC Conference on Computer and Communications Security (CCS '25), October 13--17, 2025, Taipei, Taiwan}
\acmDOI{10.1145/3719027.3765208}
\acmISBN{979-8-4007-1525-9/2025/10}

\usepackage{xspace}
\usepackage{fontawesome5}
\usepackage{makecell}
\usepackage{subcaption}

\newcommand{\artifactsUrl}{\url{https://doi.org/10.5281/zenodo.17021719}}

\newcommand{\erlangcve}{CVE-2025-32433\xspace}
\newcommand{\tectiacve}{CVE-2025-32942\xspace}

\newcommand{\sshmsg}[1]{\texttt{#1}\xspace}
\newcommand{\msgkexinit}{\sshmsg{KEXINIT}}
\newcommand{\msgnewkeys}{\sshmsg{NEWKEYS}}
\newcommand{\msgkexecdhinit}{\sshmsg{KEX\_ECDH\_INIT}}
\newcommand{\msgkexecdhreply}{\sshmsg{KEX\_ECDH\_REPLY}}
\newcommand{\msgservicerequest}{\sshmsg{SERVICE\_REQUEST}}
\newcommand{\msgserviceaccept}{\sshmsg{SERVICE\_ACCEPT}}
\newcommand{\msgignore}{\sshmsg{IGNORE}}
\newcommand{\msgdebug}{\sshmsg{DEBUG}}

\newcommand{\msgunimplemented}{\sshmsg{UNIMPLEMENTED}}
\newcommand{\msgextinfo}{\sshmsg{EXT\_INFO}}
\newcommand{\msguserauthrequest}{\sshmsg{USERAUTH\_REQUEST}}

\newcommand{\msguserauthsuccess}{\sshmsg{USERAUTH\_SUCCESS}}

\newcommand{\msgchannelopen}{\sshmsg{CHANNEL\_OPEN}}
\newcommand{\msgchannelclose}{\sshmsg{CHANNEL\_CLOSE}}
\newcommand{\msgchannelrequest}{\sshmsg{CHANNEL\_REQUEST}}
\newcommand{\msgchannelsuccess}{\sshmsg{CHANNEL\_SUCCESS}}
\newcommand{\msgchanneldata}{\sshmsg{CHANNEL\_DATA}}
\newcommand{\msgkexdhgexrequest}{\sshmsg{KEX\_DH\_GEX\_REQUEST}}

\newcommand{\supported}{\faCircle}

\newcommand{\unsupported}{\faCircle[regular]}

\definecolor{DarkGreen}{RGB}{0,152,0}

\newcommand{\nrImplementations}{ten\xspace}
\newcommand{\nrViolatingImplementationsUpper}{Seven\xspace}
\newcommand{\nrViolatingImplementations}{seven\xspace}
\newcommand{\nrStateMachines}{33\xspace}

\begin{document}

\title{Finding SSH Strict Key Exchange Violations by State Learning}

\author{Fabian Bäumer}
\orcid{0009-0006-5569-6625}
\affiliation{%
  \institution{Ruhr University Bochum}
  \city{Bochum}
  \country{Germany}
}
\email{fabian.baeumer@rub.de}

\author{Marcel Maehren}
\orcid{0009-0002-8964-826X}
\affiliation{%
  \institution{Ruhr University Bochum}
  \city{Bochum}
  \country{Germany}
}
\email{marcel.maehren@rub.de}

\author{Marcus Brinkmann}
\orcid{0000-0001-5649-6357}
\affiliation{%
  \institution{Ruhr University Bochum}
  \city{Bochum}
  \country{Germany}
}
\email{marcus.brinkmann@rub.de}

\author{Jörg Schwenk}
\orcid{0000-0001-9315-7354}
\affiliation{%
  \institution{Ruhr University Bochum}
  \city{Bochum}
  \country{Germany}
}
\email{joerg.schwenk@rub.de}

\renewcommand{\shortauthors}{Fabian Bäumer, Marcel Maehren, Marcus Brinkmann, and Jörg Schwenk}

\begin{abstract}
  SSH is an important protocol for secure remote shell access to servers on the Internet. At USENIX 2024, Bäumer et al. presented the Terrapin attack on SSH~\cite{USENIX:BauBriSch24}, which relies on the attacker injecting optional messages during the key exchange. To mitigate this attack, SSH vendors adopted an extension developed by OpenSSH called \emph{strict key exchange} (``strict KEX''). With strict KEX, optional messages are forbidden during the handshake, preventing the attack. In practice, this should simplify the state machine of an SSH handshake to a linear message flow similar to that of TLS.
  
  In this work, we analyze the design, implementation, and security of strict KEX in popular SSH servers, using black-box \emph{state learning}, which can uncover the hidden state machine of an implementation. In practice, it is limited by the number of learned messages and the complexity of the state machine. Thus, learning the \emph{complete} state machine of SSH is infeasible. Previous research on SSH, therefore, excluded optional messages, learning only a partial state machine. However, these messages are a critical part of the Terrapin attack. We propose to instead learn the complete state machine of the handshake phase of an SSH server, but with strict KEX enabled.

  We investigate the security of \nrImplementations SSH implementations supporting strict KEX for up to five key exchange algorithms. In total, we learn \nrStateMachines state machines, revealing significant differences in the implementations. We show that \nrViolatingImplementations implementations violate the strict KEX specification and find two critical security vulnerabilities. One results in a rogue session attack in the proprietary Tectia SSH implementation. Another affects the official SSH implementation of the Erlang Open Telecom Platform, and enables unauthenticated remote code execution in the security context of the SSH server.
\end{abstract}

\begin{CCSXML}
<ccs2012>
<concept>
<concept_id>10003033.10003039.10003048</concept_id>
<concept_desc>Networks~Transport protocols</concept_desc>
<concept_significance>500</concept_significance>
</concept>
<concept>
<concept_id>10003033.10003039.10003041.10003042</concept_id>
<concept_desc>Networks~Protocol testing and verification</concept_desc>
<concept_significance>500</concept_significance>
</concept>
<concept>
<concept_id>10003033.10003083.10003014.10003015</concept_id>
<concept_desc>Networks~Security protocols</concept_desc>
<concept_significance>500</concept_significance>
</concept>
</ccs2012>
\end{CCSXML}

\ccsdesc[500]{Networks~Transport protocols}
\ccsdesc[500]{Networks~Protocol testing and verification}
\ccsdesc[500]{Networks~Security protocols}

\keywords{SSH, State Learning, Strict Key Exchange, Protocol Violations, Message Injection, Man-in-the-Middle}

\maketitle

\section{Introduction}
\label{sec:intro}

\paragraph{SSH}

The Secure Shell (SSH) protocol was developed to prevent sniffing and active man-in-the-middle (MitM) attacks when an administrator executes commands on a remote server. Such servers can be stand-alone Linux or Windows systems, cloud-based virtual machines, or embedded devices such as routers, switches, and management systems. The security of SSH servers is thus of critical importance to the Internet and telecommunication networks. Since the release of the SSHv2 standard in 2006 by the IETF as a series of RFCs~\cite{rfc4250,rfc4251,rfc4252,rfc4253,rfc4254}, SSH has continuously evolved through a series of extensions, some published as separate RFCs, some developed by the community, and some proprietary to a single product. In 2024, the IETF formed the Secure Shell Maintenance~(sshm) working group, intending to revitalize the standardization of SSH under its leadership. These efforts were also motivated by the worldwide transition to post-quantum cryptography and the countermeasures against the Terrapin attack that require SSH standard changes.

\paragraph{Terrapin Attack and Strict KEX}

At USENIX 2024, Bäumer~et~al. presented the Terrapin attack~\cite{USENIX:BauBriSch24} against the integrity of the encrypted and authenticated channel in an SSH connection. The attacker can remove an arbitrary number of messages from the beginning of the channel, for example, drop an \msgextinfo message. The attack exploits that SSH allows optional messages in the handshake (e.g., the \msgignore messages in the partial state machine in \autoref{fig:ssh-hs}), and the fact that sequence numbers are not reset before switching to authenticated encryption. As mitigation, OpenSSH proposed a new protocol extension called \emph{strict KEX}~\cite{ietf-sshm-strict-kex-00}. The client and server can negotiate this protocol extension by sending a pseudo-algorithm identifier in their \msgkexinit messages. If proposed by both parties, any optional or invalid messages must cause the handshake to fail, effectively banning these messages from this phase. This restriction is also applied retroactively to any messages seen before \msgkexinit. Additionally, the two sequence number counters maintained by each peer will be reset, for each of the two unidirectional channels, after sending and receiving the \msgnewkeys message.

\begin{figure}
    \centering
    \frame{\includegraphics[width=.62\columnwidth,page=1]{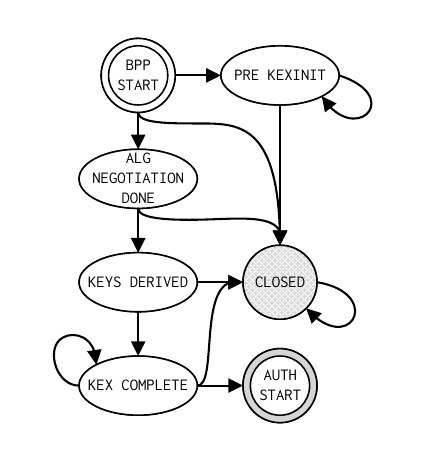}}
    \caption{A simplified state machine of a strict KEX-compliant SSH server implementation. The happy flow on the left side can only be traversed by sending the expected symbol at each protocol state. Any unexpected or unknown message causes the connection to terminate immediately. \texttt{PRE KEXINIT} represents the subgraph before sending \msgkexinit.}
    \Description{The figure depicts a simplified state machine of an SSH server that has strict KEX enabled. The state machine starts in the ``BBP START'' state and transitions through the happy flow on the left consisting of the states ``ALG NEGOTIATION DONE'', ``KEYS DERIVED'', ``KEX COMPLETE'', and the accepting state ``AUTH START''. All unexpected messages cause the connection to terminate, which is also the case for messages sent before the KEXINIT messages.}
    \label{fig:ssh-strict-kex-sm}
\end{figure}

A secure implementation requires both the sequence number reset and strict message handling to be done correctly. As any violation of the sequence number reset would prevent a peer from using the secure channel, we assume it is correctly implemented. For strict message handling, the implementation's state machine of the key exchange must be modified; a compliant server state machine is depicted in \autoref{fig:ssh-strict-kex-sm}. At any point during the handshake after \msgkexinit, a peer receiving optional or invalid messages must terminate the connection. However, peers may allow optional messages before \msgkexinit, but must consequently terminate the connection if any such message is received and strict KEX is negotiated.

Such changes to an existing implementation can introduce new errors due to the complexity of the SSH protocol. Developers might make mistakes that are difficult to find because a non-compliant implementation can still be fully functional in development, testing, and deployment. Finding a violation amounts to the problem of finding \emph{any} message flow that does not conform to the specification, resulting in an exponential search space. Thus, we require an automated approach to find such violations.

\paragraph{State Machine Learning (SML)}

SML is a technique where a \emph{learner} uses a learning algorithm (for example, $L^*$~\cite{angluin1987learning} or TTT~\cite{isberner2014ttt}) to infer a hypothesis about a state machine by interaction. The learning algorithm uses a fixed \emph{alphabet} consisting of \emph{symbols} to formulate membership queries. In practice, a \emph{mapper} must translate these symbols into concrete inputs that the \emph{system under learning} (SUL) can understand and respond to. With each new hypothesis, an \emph{equivalence oracle} is queried to find counterexamples disproving the hypothesis. Any counterexample found can then be passed to the learning algorithm to refine the hypothesis.

The size and degree of detail of the retrieved Mealy automaton may differ significantly: It depends on the abstraction level of the learner and the feedback it gets from the \emph{system under learning} (SUL). For example, if we can observe the CPU and its registers, we can easily distinguish exponentially many CPU states with limited insight. If, on the other hand, we can only distinguish different responses sent from the SUL over a network, then we will only learn a much simpler Mealy automaton---but this automaton will give us valuable insights into the logical structure of the SUL. Typical security errors found with this latter configuration are authentication bypasses, where an attacker can bypass client authentication by sending an unexpected, non-standard sequence of messages~\cite{USENIX:FJMRSS20}.

\paragraph{SML on SSH}

Inferring state machines from SSH implementations is challenging due to optional messages, which enable arbitrarily long protocol flows even before the handshake is completed. For example, all of these flows end up in \texttt{KEX COMPLETE} of \autoref{fig:ssh-hs}:

\indent {\small (\msgkexinit, \sshmsg{ECDH\_INIT}, \msgnewkeys, \msgservicerequest),}\\ 
\indent {\small (\msgkexinit, \msgignore, \sshmsg{ECDH\_INIT}, \msgnewkeys, \msgservicerequest),}\\
\indent {\small (\msgkexinit, \msgignore, \msgignore, \sshmsg{ECDH\_INIT}, \msgnewkeys, \msgservicerequest),}\\
For this reason, previous research~\cite{SPIN:fiterauSshModelChecking,tijssen2014automatic,NDSS:FJST23} did not include optional messages in their alphabet, effectively learning only a partial model of the implementation's state machine. However, optional messages were instrumental in the Terrapin attack, so such simplifications could hide details relevant to the security of the protocol.

\newcommand{\adapted}{\textsuperscript{\textdagger}}
\begin{table*}
    \centering
    \small
    \caption{Comparison of research publications on active state learning of SSH and other security-related protocols, sorted by publication year and author names. We provide the protocol of the system under learning, the number of learned state machines, the number of symbols in the input alphabet, and details about the mapper and equivalence oracle. A dagger symbol (\textdagger) indicates that the mapper implementation is modified to enable state learning. If multiple alphabets are used during learning, we give an upper bound for the number of symbols in the alphabet.}
    \label{tab:related}
    \begin{tabular}{lcclrrll}
        \toprule
        Authors & Year & Ref. & Protocol & \# SMs & \# Symbols & Mapper & Eq. Oracle \\
        \midrule
        de Ruiter~et~al. & 2015 & \cite{USENIX:dRuPol15} & TLS & 16 & $\leq 13$ & \emph{novel} & W-method, modified  \\
        de Ruiter & 2016 & \cite{SecITSys:dRu16} & TLS & 290 & $\leq 8$ & \multicolumn{2}{c}{\cellcolor{black!10} Reuses learner from~\cite{USENIX:dRuPol15}} \\
        Fiter\u{a}u-Bro\c{s}tean~et~al. & 2017 & \cite{SPIN:fiterauSshModelChecking} & SSH & 3 & $\leq 21$ & Paramiko\adapted & \cite{automatalearning-ecs} \\
        Daniel~et~al. & 2018 & \cite{danielopenvpnstatelearning} & OpenVPN & 12 & $\leq 12$ & based on de Ruiter~et~al.~\cite{USENIX:dRuPol15} & W-method, modified \\
        Stone~et~al. & 2018 & \cite{ESORICS:StoChodRu18} & 802.11 & 7 & 45 & Scapy & \emph{novel} + W-method \\
        Yadav~et~al. & 2018 & \cite{SSCC:YadSad19} & TLS & 6 & \emph{unknown} & \emph{unknown} & \emph{unknown} \\
        Guo~et~al. & 2019 & \cite{8913552} & IPSec & 3 & 17 & Scapy & \cite{533956} \\
        Rasool~et~al. & 2019 & \cite{rasool2019state} & QUIC & 2 & $\leq 5$ & \emph{novel} & random words \\
        Fiter\u{a}u-Bro\c{s}tean~et~al. & 2020 & \cite{USENIX:FJMRSS20} & DTLS & 31 & 17 & TLS-Attacker~\cite{2024TLSAttacker} & (random) Wp-method \\
        Chlosta~et~al. & 2021 & \cite{WISEC:ChlRupHol21} & LTE & 7 & $\leq 67$ & srsRAN\adapted & random words \\
        Ferreira~et~al. & 2021 & \cite{10.1145/3452296.3472938} & QUIC & 3 & 7 & \emph{novel} & Wp-method \\
        Hussain~et~al. & 2021 & \cite{CCS:HKICB21} & LTE & 14 & 35 & srsLTE\adapted & \emph{novel} \\
        Pferscher~et~al. & 2021 & \cite{FM:PfeAic21} & BLE & 7 & $\leq 9$ & Scapy & state prefixed random walk \\
        Pferscher~et~al. & 2022 & \cite{NFM:PfeAic22} & BLE & 8 & $\leq 9$ & \multicolumn{2}{c}{\cellcolor{black!10} Reuses learner from~\cite{FM:PfeAic21}} \\
        Rasoamanana~et~al. & 2022 & \cite{ESORICS:RasLevDeb22} & TLS & $> 2000$ & \emph{unknown} & TLS-Attacker~\cite{2024TLSAttacker} & random walk \\
        Stone~et~al. & 2022 & \cite{CCS:STVHBC22} & TLS & 14 & \emph{unknown} & based on de Ruiter~et~al.~\cite{USENIX:dRuPol15} & W-method, modified \\
        ~ & ~ & ~ & 802.11 & 2 & $\leq 40$ & based on Stone~et~al.~\cite{ESORICS:StoChodRu18} & W-method, modified \\
        Fiter\u{a}u-Bro\c{s}tean~et~al. & 2023 & \cite{NDSS:FJST23} & SSH & 3 & 21 & \multicolumn{2}{c}{\cellcolor{black!10} Reuses learner from~\cite{SPIN:fiterauSshModelChecking}}  \\
        ~ & ~ & ~ & DTLS & 16 & 17 & \multicolumn{2}{c}{\cellcolor{black!10} Reuses learner from~\cite{USENIX:FJMRSS20}} \\
        Karim~et~al. & 2023 & \cite{SP:KIHB23} & BLE & 25 & 32 & SweynTooth\adapted~\cite{sweyntooth} & Wp-method \\
        Tu~et~al. & 2024 & \cite{USENIX:TIRDWW24} & 5G CPP & 19 & 29 & OpenAirInterface\adapted / srsRAN\adapted / Open5GS\adapted & Wp-method  \\
        Dong~et~al. & 2025 & \cite{dong2025corecrisis} & 5G Core & 4 & $18 + x$ & UERANSIM\adapted & \emph{novel} \\ 
        Ang~et~al. & 2025 & \cite{quic-states-2025} & QUIC & 186 & 26 &  aioquic\adapted & Wp-method \\
        Maehren~et~al. & 2025 & \cite{maehren2025towards} & TLS & 1304 & 60 & TLS-Attacker~\cite{2024TLSAttacker} & \emph{novel} \\ 
        \midrule
        This work & ~ & ~ & SSH & \nrStateMachines & 94 & \emph{novel} & \emph{novel} + random words \\
        \bottomrule
    \end{tabular}
\end{table*}

In \autoref{tab:related}, we provide a comparison of prior works applying SML to various protocols. Besides excluding invalid messages, two previous studies focusing on SSH employed a relatively limited alphabet consisting of at most 21 input symbols. To achieve a thorough but practical analysis of the SSH key exchange extending beyond the limited alphabets of previous works, we adopted state learning to SSH by the following strategy: (i) Alphabet: Our novel contribution is the largest alphabet for SSH used to date, which is adapted to the analysis of strict KEX. Additionally, we use ``forbidden'' symbols during the handshake, which enabled us to find a critical Erlang vulnerability. (ii) Mapper: Using our own custom SSH implementation, we introduce a novel mapper that can create arbitrary valid or invalid SSH messages at any time during the connection. These messages require dynamic generation based on the cryptographic connection state. (iii) Equivalence Oracle: A generic equivalence oracle leads to overlong learning times (see \autoref{sec:eval:sub:equiv}), with little coverage of the state machine. We propose an oracle that enables TTT to find unexpected state transitions starting from deep paths within a reasonable runtime, utilizing a randomized component.

\subsection{Research Questions}

With strict KEX enabled, all optional messages \emph{should} terminate the connection. However, as this requirement is not cryptographically ensured, the question remains whether this is the case in practice.

\begin{quote}
    RQ 1: Do implementations of strict KEX correctly ban all optional or invalid messages in all edge cases?
\end{quote}
We answer this question in the negative. In fact, we find that implementers used a variety of strategies to implement strict KEX, but many were not successful in completely implementing it. \nrViolatingImplementationsUpper out of \nrImplementations implementations do not implement strict KEX correctly.

\paragraph{Security Vulnerabilities in Strict KEX Implementations}

After finding non-compliant implementations of SSH strict KEX, we extend our research question:
\begin{quote}
    RQ 2: Do violations of the strict KEX specification lead to exploitable security vulnerabilities?
\end{quote}
We answer this question affirmatively. We find two implementations that are vulnerable to powerful practical attacks, compromising the security of the client or server. The first attack is against Tectia SSH, a proprietary implementation by a company founded by Tatu Ylönen (the inventor of the SSH protocol). It is a rogue session attack against the client, where a MitM attacker with a shell account on the vulnerable server can log the victim user into the attacker's account. Henceforth, the attacker can control the terminal output seen by the user and extract the user's keyboard input, allowing for phishing attacks. The second attack affects the official SSH implementation of the Erlang Open Telecom Platform. It is a pre-authentication remote code execution that allows an attacker with network access to the SSH server to execute arbitrary code in the security context of the SSH daemon.

\paragraph{Contributions}

Our main contributions are as follows:
\begin{itemize}
    \item We provide the first security analysis of the SSH strict KEX countermeasure to the Terrapin attack in SSH servers, analyzing \nrImplementations implementations across five different key exchange flows. We find that \nrViolatingImplementations implementations fail to implement strict KEX correctly, allowing optional or invalid messages during the handshake, despite strict KEX being active.

    \item We provide an implementation of a state learner for the SSH protocol tailored to the analysis of the key exchange phase with an extensive alphabet exploring all defined and selected undefined message types, allowing for an efficient and exhaustive analysis of this crucial phase of the protocol.

    \item We provide \nrStateMachines learned state machines for \nrImplementations SSH server implementations, supporting up to five different key exchange flow types, with strict KEX enabled.

    \item We demonstrate critical security vulnerabilities in two SSH servers, allowing a MitM attacker to launch a rogue session attack against another user in one server and execute remote code before authentication in the other.
\end{itemize}

\paragraph{Artifacts}

The artifacts contain our custom SSH implementation written in Java, our SSH state learner written in Kotlin, proofs-of-concept for the attacks against Tectia SSH (\autoref{sec:eval:sub:tectia}) and Erlang SSH (\autoref{sec:eval:sub:erlang}), as well as the state machines that we obtained during evaluation. We also include one protocol flow definition violating strict KEX requirements per affected implementation and additional material. 
The artifacts are available at:
\begin{quote}
    \artifactsUrl
\end{quote}

\paragraph{Ethical Considerations}

We reveal several violations against strict KEX, a security-relevant protocol extension of the SSH protocol. While not all violations pose an immediate security threat to users, this may change as the protocol evolves. Additionally, we identify critical vulnerabilities in Tectia SSH and Erlang SSH that can be exploited in practice. We initiated a responsible disclosure process with each affected vendor. The Erlang developers have confirmed our findings and provided an updated version as part of a public disclosure. Cisco\footnote{\url{https://sec.cloudapps.cisco.com/security/center/content/CiscoSecurityAdvisory/cisco-sa-erlang-otp-ssh-xyZZy}, accessed: 2025-09-09} and NetApp\footnote{\url{https://security.netapp.com/advisory/ntap-20250425-0001/}, accessed: 2025-09-09} published a list of affected devices. The vendor of Tectia SSH also confirmed our findings and distributed an updated version to their customers. The public disclosure for this vulnerability is currently scheduled for mid-September. We were awarded a €$2{,}000$ bug bounty for the Tectia SSH vulnerability.

\section{Background}

\subsection{SSH}

\begin{figure}[t]
    \centering
    \frame{\includegraphics[width=\columnwidth]{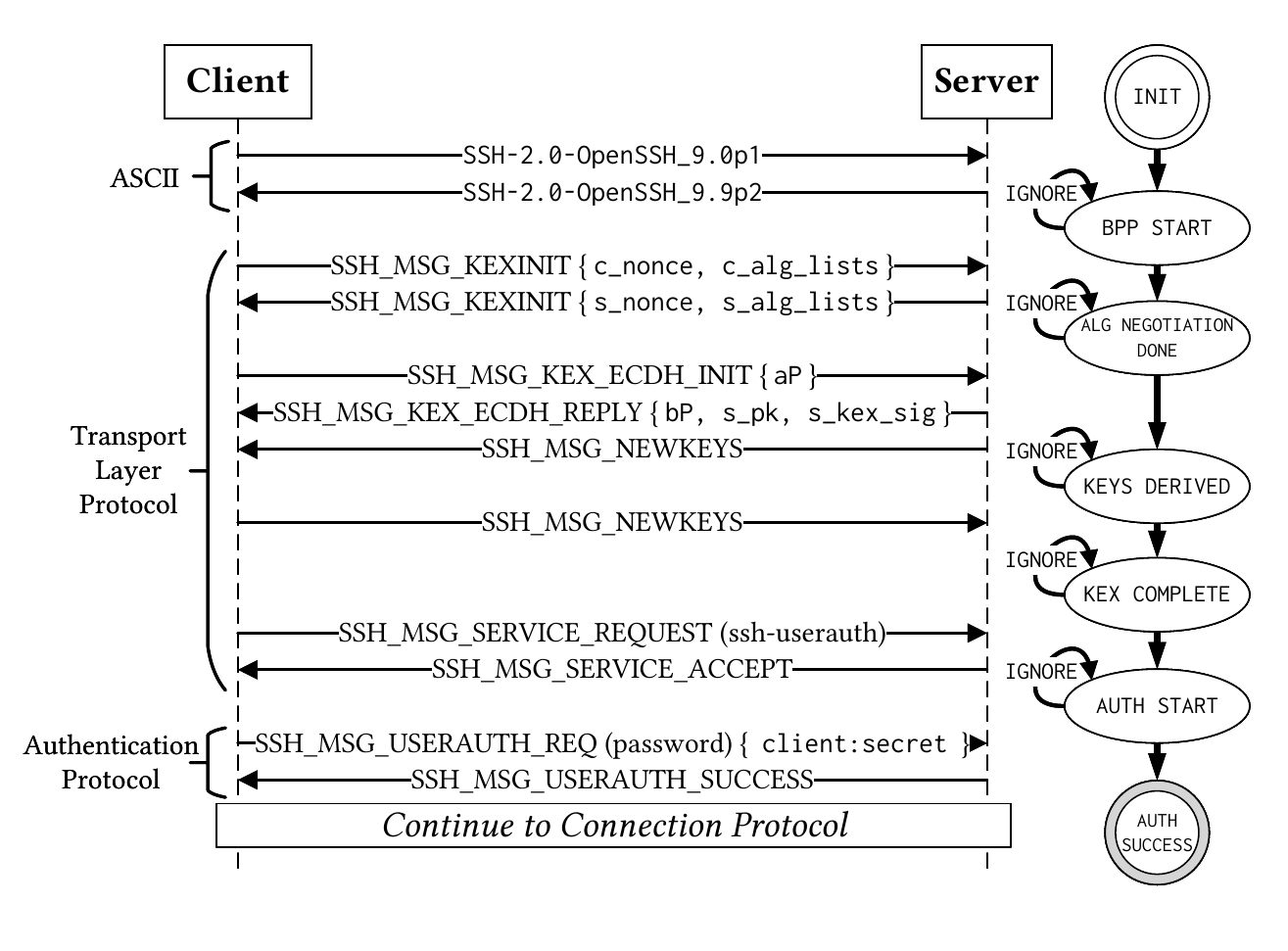}}
    \caption{SSH handshake example, with the corresponding (partial) state machine of the server. State transitions are caused by the interleaved incoming messages, and their output is shown by the interleaved outgoing messages. We analyze this state machine for strict KEX violations. Note that the message flow on the left only shows mandatory messages, but SSH allows for various optional messages, e.g., the \msgignore messages included in the partial state machine.}
    \Description{The figure depicts a typical SSH protocol handshake using an elliptic curve Diffie-Hellman key exchange. Client and server start by sending ASCII banners with their supported SSH version and software version, followed by algorithm negotiation via the \msgkexinit messages. The client then initiates the actual key exchange by sending \msgkexecdhinit, to which the server responds with \msgkexecdhreply and the \msgnewkeys message. The message exchange then continues inside a secure channel where the client requests the user authentication service and performs user authentication. On the right side, the corresponding state diagram of the server is depicted, which transitions according to the messages sent and received.}
    \label{fig:ssh-hs}
\end{figure}

\paragraph{SSH Protocol} 

An example of an SSH key exchange is shown in \autoref{fig:ssh-hs}. First, an ASCII banner is exchanged, followed by a \msgkexinit message containing algorithm proposals and nonces. Then, the algorithm-specific key exchange protocol is executed, which in the case of ECDH consists of the \msgkexecdhinit message, with the client's ephemeral public key, followed by a \msgkexecdhreply message containing the server's ephemeral public key and host key signature. The signature authenticates the server and critical parts of the key exchange so far. Finally, both parties send a \msgnewkeys message to complete the handshake and activate encryption. Both parties continue with the user authentication protocol. The client sends \msguserauthrequest messages to authenticate itself, e.g., with a username and password. If these are correct, the server will send the \msguserauthsuccess message. Afterward, both parties switch to the connection protocol, which provides the remote shell.

\paragraph{Optional Messages} 

The SSH specification defines message types that either party may optionally send. For example, the \msgignore message can be sent at any time to thwart traffic analysis attacks~\cite{USENIX:SonWagTia01}. Another example is the \msgextinfo message, which can be used to negotiate protocol extensions securely within the secure channel.

\paragraph{Sequence Numbers in the Binary Packet Protocol} 

SSH uses sequence numbers (SQNs) to prevent man-in-the-middle attackers from dropping or re-arranging messages conveyed through the secure channel by appending a MAC over the SQN to each packet. SQNs are implemented as two counters, one for each connection direction: a send counter and a receive counter. Unless an attacker manipulates the message sequence, the send counter of one end and the receive counter of the other end should always align. Changing the order of a message results in an invalid MAC, as the recipient would use an incorrect SQN during validation. Similarly, dropping a message would invalidate the MAC of the following message on the recipient's end. Note that unprotected messages sent before the \msgnewkeys message already increment the SQNs, but SQN verification only begins afterward, as there is no MAC included earlier.

\paragraph{Terrapin Attack and Strict KEX} 

The Terrapin attack presented by Bäumer~et~al.~\cite{USENIX:BauBriSch24} is a prefix truncation attack, allowing an attacker to delete initial messages from the secure channel. The attack exploits that some cipher modes continue to use the SQNs of the handshake inside the secure channel, in combination with the protocol's tolerance for optional messages. To delete a message, an attacker would first inject an optional message, such as \msgignore, which increments the receive counter of the recipient and misaligns it with the send counter of its peer. Then, after both parties sent their respective \msgnewkeys messages, the attacker can drop the first message sent to this peer, thus realigning the receive counter with the send counter of the other party. Consequently, future protected messages will not reveal that the attacker deleted a message from the secure channel, as the MAC validation succeeds.

To mitigate the attack, the server and client can negotiate a \emph{strict KEX} mode that forces sequence numbers to be reset to zero when cryptographic key material is activated and defines any optional messages sent prior to \msgnewkeys as invalid, rendering the SQN manipulation technique of the Terrapin attack impossible. However, the initial definition of this requirement is quite informal~\cite[Section 1.10]{protocolopenssh}: \emph{``During initial KEX, terminate the connection if an out-of-sequence packet or any message that is not strictly required by KEX is received. This includes terminating the connection if the first packet received is not SSH2\_MSG\_KEXINIT. Unexpected packets in the context of strict KEX include messages that are otherwise  valid at any time during the connection such as SSH2\_MSG\_DEBUG, SSH2\_MSG\_IGNORE or SSH2\_MSG\_UNIMPLEMENTED.''} A more formal specification~\cite{ietf-sshm-strict-kex-00} of this countermeasure is currently being discussed for adoption by the sshm IETF working group.

\subsection{State Learning}

State Learning usually consists of four elements: a \textit{learner}, a \textit{mapper}, an \textit{equivalence oracle}, and a \textit{system under learning} (SUL). The user defines an abstract set of input symbols (\textit{alphabet}). The learner constructs chains of symbols (\textit{words}) to formulate queries. The mapper receives these queries and transforms the individual abstract symbols into meaningful protocol messages to be sent to the SUL. The mapper collects the responses of the SUL, transforms them into abstract protocol messages, and returns them to the learner as the result of the requested query. Providing an abstraction for the responses allows the learner to discern semantic differences in the output, rather than meaningless differences caused, for example, by random nonces in each invocation of the protocol. By observing the output of different queries, the learner generates a \textit{hypothesis} of the SUL's state machine. Ideally, this hypothesis can be used to accurately predict the output of the SUL for any sequence of symbols of the alphabet.

\paragraph{Equivalence Tests}

After forming a hypothesis, the \textit{equivalence oracle} checks its correctness by generating queries and comparing the responses of the hypothesis and the SUL. A mismatch yields a \textit{counterexample}, which the learner uses to refine the hypothesis. This process is repeated until no more counterexamples are found. The final hypothesis is then assumed to represent the SUL accurately. Various strategies, such as random queries, are used to find counterexamples.

\paragraph{Choosing the Alphabet}

Designing the input alphabet is a key challenge in state learning. It requires expert knowledge of the protocol to define symbols that enable the learner to distinguish states based on observable responses. Symbols that do not reveal behavioral differences should be avoided, as execution time typically grows exponentially with alphabet size. Since the actual effect of each symbol becomes apparent only during learning, input selection must rely on informed assumptions.

\section{Methodology}

Our study aims to find bypasses to the handshake hardening measures in the strict KEX implementations \emph{of SSH servers}. Specifically, we aim to identify messages that can be injected into the happy flow of messages sent to the server without causing immediate session termination, as mandated by strict KEX. We chose state machine learning (SML) as our main methodology, as it offers a systematic approach to protocol implementation analysis and is well-suited to uncover such violations. If an implementation tolerates unexpected messages, this should manifest as a self-loop in the structure of the inferred state machine. SML has been used before to analyze cryptographic protocol implementations, and authentication bypasses were found with SML in the case of DTLS~\cite{USENIX:FJMRSS20}.

\subsection{Architecture}
\label{sec:methodology:sub:architecture}

\begin{figure}
    \centering
    \includegraphics[width=0.8\linewidth]{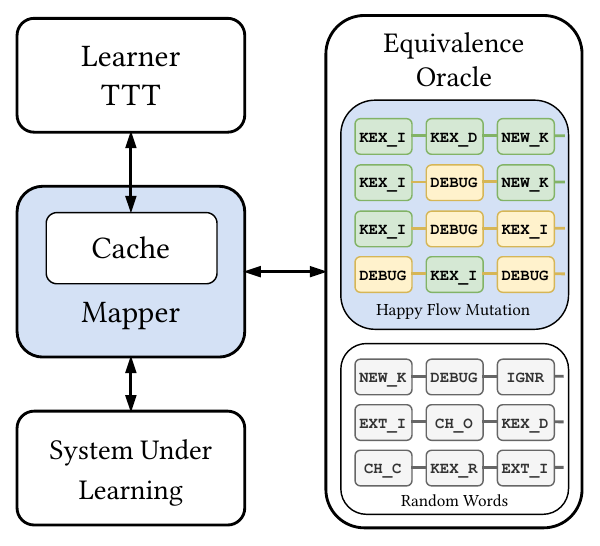}
    \caption{Illustration of the components of our state learner. New contributions are highlighted in blue. Our happy flow mutation strategy for the equivalence oracle injects misplaced messages (yellow) into the chain of valid messages (green). In contrast, the supplementary random words approach, often used for state learning, chooses random queries regardless of the validity of the created flow.}
    \Description{The figure describes the different components of our state learner. The components include the learner implementing the TTT algorithm, the mapper with a cache, the system under learning, and the equivalence oracle. The mapper is the connecting component of the three other components. For the equivalence oracle, happy flow mutations and random words are depicted.}
    \label{fig:state-learner-overview}
\end{figure}

\autoref{fig:state-learner-overview} depicts the architecture used in our evaluation. The \emph{learner} uses the TTT algorithm---more precisely, the LearnLib library~\cite{isberner2014ttt,isberner2015open} by Isberner, Howar, and Steffen---which issues conjectures about the state machine of the \emph{system under learning} (SUL) based on responses from the SUL to inputs taken from the learning alphabet. In our case, the SULs are different SSH server implementations. Learner and SUL speak different languages: The learner only outputs and understands abstract symbols, while the SUL requires valid SSH client messages and responds with valid SSH server messages. The \emph{mapper} translates between the two entities. One of the design choices to be made in SML is how to map abstract symbols to SSH messages and vice versa, as detailed in subsections~\ref{subsec:alphabet} and~\ref{subsec:mapper}.

The learner assumes that the SUL is deterministic; that is, the system always responds with the same answers when queried with the same message sequences. When the learner queries the same sequence of symbols repeatedly, we already know the answers of the SUL from the first run-through of this sequence. As in previous works~\cite{SP:SAPKJ17,CCS:ASJKK16,SP:KIHB23}, we can always answer consecutive queries for the same sequence of symbols from a \emph{cache} to enhance performance. However, as soon as a single symbol is added to a known sequence that does not already end in a closed connection state, and this extended sequence has never been asked before, we cannot rely on the cache. We have to invoke the SUL again, replaying the common prefix to reproduce the SUL's internal state, before we can answer the additional symbol. As discussed in \autoref{par:nondeterminism}, we may observe non-determinism due to internal timings and inconsistent behavior. In this case, the response to a replayed prefix conflicts with the cache. In such cases, we use high-confidence majority voting to resolve the dispute.

After some queries via the mapper to the SUL, the learner will issue a \emph{hypothesis} on the state machine of the SUL. In the beginning, this hypothesis may only consist of a single state. To challenge this hypothesis and to force the learner to refine it, an \emph{equivalence oracle} is invoked that outputs several sequences of symbols. These \emph{equivalence tests} are used to, on the one hand, traverse the state machine hypothesis issued by the learner and to invoke the actual SUL through the mapper. If the outputs of the hypothesis and the SUL are always identical, the hypothesis is confirmed, and the SML terminates. If the outputs differ, the current hypothesis is disproven as it does not accurately represent the state machine of the SUL, and the learner then uses this counterexample to refine the hypothesis.

A generic equivalence oracle outputs random sequences of symbols from the alphabet; we use such an oracle implemented by the LearnLib library. A random words oracle closely resembles random binary fuzzing, but at a different level of abstraction. Like random binary fuzzing, this is the best strategy to find all deviations in an implementation, but it may consume an infeasible amount of time. Therefore, we add a second equivalence oracle, \emph{happy flow mutation}, which directly targets violations of strict KEX implementations (see \autoref{subsec:eqoracle}). The mapper and happy flow mutation components form the novel tooling support for our study.

\subsection{Alphabet} 
\label{subsec:alphabet}

To analyze if servers tolerate any misplaced messages, we include \emph{all} SSH client messages from the transport layer~\cite{rfc4253}, authentication~\cite{rfc4252}, and connection protocol~\cite{rfc4254}. These are defined either in the SSH RFCs mentioned above or in additional RFCs~\cite{rfc8308, rfc4256, rfc4419, rfc4335, rfc4432, rfc5656}, and we also include messages from well-known vendor extensions of OpenSSH. In addition to those messages sent from the client to the server, we include messages usually expected to be sent exclusively by the server, such as \msgkexecdhreply. Furthermore, we include error-related messages, such as the \msgunimplemented and optional \msgignore and \msgextinfo messages. 

Lastly, we add symbols for undefined messages, as these might reveal corner cases of the message handling. The SSH protocol distinguishes messages based on a one-byte message number. The IANA registry for SSH parameters further partitions the byte range into multiple blocks for the transport layer protocol, the authentication protocol, and the connection protocol. We add one symbol translating to a message number in each of the ten unassigned message blocks. Specifically, we construct 1-byte messages solely consisting of the message type, which corresponds to the structure of other signaling messages of SSH, such as \msgnewkeys. 

In total, our alphabet consists of 94 symbols. The complete list of included symbols, with their corresponding SSH message type and specification reference, is provided in the artifacts.

\subsection{Mapper} 
\label{subsec:mapper}

\paragraph{Message Directions}

Our mapper strictly maps the alphabet symbols to \emph{valid} SSH messages. Some messages, such as \msgkexecdhinit and \msgkexecdhreply, require complex cryptographic computations to be valid. Additionally, to send \msgservicerequest in the encrypted channel, all SSH keys must be derived. As mentioned previously, the mapper is not restricted to messages sent from the client to the server. Consequently, the mapper contains our implementation of a standard-compliant SSH client and server; when queried, it maintains a state that allows it to answer cryptographic messages sent later in the handshake.

\paragraph{Final State}

In our experiment, we only investigate the beginning of the server's SSH state machine. The target state of this partial state machine is the state where the SUL has received and accepted the \msgservicerequest message. To distinguish different states reached by \msgservicerequest messages, we allow the learner to continue into the authentication protocol. To exclude states of the connection protocol, the mapper uses an incorrect username-password combination not known to the SUL in order to fail user authentication. Therefore, any \msguserauthsuccess message returned would directly indicate an authentication bypass.

\paragraph{Message Contents}

We distinguish three classes of messages:
\begin{itemize}
    \item \textbf{Static.} Message types like \msgnewkeys contain static content. There is a one-to-one correspondence between these message types and the actual messages.
    \item \textbf{Cryptographic.} Message types like \msgkexecdhinit and \msgkexecdhreply contain cryptographic values computed based on the states of the mapper and the SUL.
    \item \textbf{Variable.} Different options exist for the contents of other messages, which may reveal different details of the SUL in the learned state machine. A prominent example of this class of messages is the \msgkexinit message, which includes lists of supported algorithms. These lists---covering key exchange, signature, encryption, MAC, and compression---may each include an arbitrary subset of supported algorithms, resulting in an exponentially large number of possible contents.
\end{itemize}

While our mapper can deterministically form \emph{static} and \emph{cryptographic} messages, and thus provide a one-to-one mapping from the message type to the actual message itself, this is not possible for \emph{variable} messages. For \msgkexinit, we consider the following key exchange flow types:
\begin{itemize}
    \item \textbf{ECDH.} Algorithms of this flow type perform an elliptic curve Diffie-Hellman key exchange on a named curve. The protocol flow will resemble \autoref{fig:ssh-hs}.
    \item \textbf{DH.} Similar to ECDH, but a multiplicative group of integers modulo $p$ will be used instead.
    \item \textbf{DHGEX.} Here, an additional message exchange selecting the Diffie-Hellman group is introduced. The actual key exchange is identical to the previous one.
    \item \textbf{RSA.} If this (seldomly supported) key exchange flow type is selected through the \msgkexinit messages, the client generates a shared secret and encrypts it using a previously exchanged RSA public key from the server.
    \item \textbf{PQ-Hybrid.} This flow type combines a classical key exchange with a post-quantum algorithm, and the protocol flow is similar to ECDH.
\end{itemize}
Since these key exchange flow types change the state machine activated within the SUL, we learn \emph{up to five different state machines for each SUL}: For each supported key exchange flow type of the SUL (see \autoref{tab:eval_ssh_impl}), we included one algorithm of the corresponding flow type in the \msgkexinit message from the mapper to the SUL, forcing the SUL to activate the corresponding state machine. This selection is made through a configuration in the mapper; the TTT algorithm and the random words equivalence oracle remain unaffected, but the happy flow mutation oracle must be adapted to this choice.

We assume that other \msgkexinit parameters, like encryption and MAC algorithms, are unlikely to affect the state machine. Since a thorough study of the side effects of such choices would require a different methodology, it is out of scope for this work. The exact mapping of key exchange flow types to actual algorithms and our choices for all other algorithm lists are described in the artifacts.

\subsection{Equivalence Oracle} 
\label{subsec:eqoracle}

To find counterexamples, we apply a novel strategy tailored to our research question, which we call a \emph{happy flow mutation} oracle. This oracle follows the following algorithm:
\begin{enumerate}
    \item Construct the happy flow $\mathcal{F}$ for the chosen key exchange flow type, and represent it as a sequence of symbols.
    \item Pick up to $n$ symbols from the alphabet $\mathcal{A}$  (with repetitions).
    \item Choose a corresponding number of positions before, after, or within the happy flow symbol sequence.
    \item Insert the chosen symbol(s) at the chosen position(s).
\end{enumerate}
The oracle runs this algorithm exhaustively, generating every possible combination for up to $n$ symbol insertions into the happy flow. This results in a total of \[\sum_{i = 0}^n |\mathcal{A}|^i\cdot\binom{|\mathcal{F}| + i}{i} \] sequences, where $|\mathcal{A}|$ is the size of the alphabet and $|\mathcal{F}|$ is the length of the happy flow. We choose $n = 2$ with an alphabet size of $|\mathcal{A}| = 94$, resulting in $133{,}011$ sequences ($|\mathcal{F}| = 4$), or $186{,}121$ sequences for DHGEX ($|\mathcal{F}| = 5$). Effectively, this yields a thorough analysis of specific parts of a hypothesis while omitting parts of the state machine that are unlikely to result in insights relevant to our research question. Our approach guarantees to reveal tolerances for any misplaced message, while allowing for a large alphabet.

Note that the learner cannot succeed in client authentication as per the definition of the authentication request symbols in our alphabet. Therefore, we only include messages from the connection protocol to test their effect when sent during the handshake.

We supplement our happy flow manipulation equivalence oracle with a generic random words approach, generating chains of randomly sampled alphabet symbols. By doing so, we also gain limited insights into states existing further from the happy flow. The interaction of our equivalence oracle with the other components of state learning is shown in \autoref{fig:state-learner-overview}. We run the random words strategy 10,000 times using sequences ranging from 5 to 15 symbols. If neither strategy yields a counterexample, the hypothesis is considered final, thereby concluding the learning process.

\subsection{Implementation}

\paragraph{Dynamic SSH Stack and State Learner}

To conduct state learning, the mapper must be able to find suitable translations for the abstract symbols queried by the learner. This can be challenging, as message fields often depend on other messages. When sending a message outside the protocol-conformant protocol flow, the required session parameters may not yet be established. Hence, a dynamic SSH stack is needed that generates protocol-compliant messages where possible, and otherwise falls back to reasonable default values. Furthermore, some input symbols may be intended to translate to invalid messages; for example, we include symbols for undefined protocol messages in our alphabet. The underlying SSH library must provide an API to control the content of these messages. To address these challenges, we have developed our own SSH implementation, drawing on design principles from the well-known TLS-Attacker~\cite{CCS:Somorovsky16} research suite, which has been utilized in various studies~\cite[Table 1]{2024TLSAttacker}.

The state learner is written in Kotlin and uses LearnLib~\cite{isberner2015open}, a framework for state learning. We use LearnLib's implementation of the TTT algorithm as our learner and employ their random words equivalence oracle to supplement our happy flow mutation oracle.

\paragraph{Mapper} 

To prevent parsing-related side effects, we send the undefined message types in our alphabet as empty messages consisting solely of the type field. We initialize a new cipher state and reset the SQN counter for the corresponding direction whenever a \msgnewkeys message is sent to or received from the SUL. If algorithms or key material are unavailable, the mapper defaults to a \texttt{none} cipher without any MAC algorithm. Whenever we prepare a message, including those usually only sent by servers, we update our context accordingly, assuming the peer handles the message accordingly.

\subsection{Restrictions}

\paragraph{Static ASCII Banner}

Some implementations may use the ASCII banner sent at the beginning of the SSH handshake (\autoref{fig:ssh-hs}) to activate compatibility measures, influencing the state machine of the SUL by changing the behavior of the SSH server. We did not investigate this in our experiment, and set the client banner to the static value \texttt{SSH-2.0-OpenSSH\_9.0}.

\paragraph{Invalid Messages}

We do not send cryptographically or structurally invalid messages to the server. That is, all messages are constructed with valid cryptographic values and correct structural components, such as length fields and padding. Messages that represent unknown messages consist of a single byte containing the message ID.

\paragraph{Static KEXINIT Symbol}

We do not use multiple \msgkexinit symbols with different algorithm lists for a single learner run. Instead, the list of supported key exchange algorithms is populated with an algorithm that is provided to the learner at invocation. For all other algorithm lists, we choose reasonable defaults based on the supported algorithms of OpenSSH.

\subsection{Post Analysis}
\label{sub:postAnalysis}

After extracting the state machine, we iterate over the states of the happy flow, testing for inputs to the transition function that are forbidden under strict KEX, which do not cause the handshake to fail. In particular, we search for invalid messages sent after strict KEX has been negotiated, as well as messages sent beforehand that remain unnoticed and do not cause the session to fail retrospectively once strict KEX is in effect. For any such deviation from the strict KEX specification, we manually implement a dedicated test flow using our SSH implementation to confirm and further investigate the impact of the identified issues. To correctly detect taint-based implementations, each test flow performs a complete run of the transport layer protocol, eventually requesting user authentication after algorithm negotiation and key exchange. If the server responds to this request by accepting the service despite the strict KEX violation present, we consider the strict KEX violation confirmed.

\section{Evaluation}
\label{sec:eval}

We select \nrImplementations SSH server implementations that, according to the official Terrapin attack website,\footnote{\url{https://terrapin-attack.com/patches.html}, accessed: 2025-04-14} implement the strict KEX countermeasure (see \autoref{tab:eval_ssh_impl}). We only tested the latest version of each implementation to exclude early versions of the countermeasure with potentially faulty implementations, since strict KEX was introduced less than two years ago. Our selection includes well-known servers like OpenSSH, Dropbear, Bitvise SSH, and Tectia SSH, and libraries such as AsyncSSH, Golang's x/crypto/ssh, and libssh, which is used by GitLab and by KDE (a desktop environment for Linux). LANCOM LCOS is part of the firmware of an embedded network device, and Erlang SSH is part of OTP, the Open Telecom Platform by Ericsson, commonly used in network devices. TinySSH is included because it is a minimalist server implementation focusing on modern cryptographic primitives, such as post-quantum algorithms. All implementations support the ECDH key exchange flow, and most also support DH and DHGEX, while six support the selected PQ-hybrid key exchange. Only one implementation supports RSA. We evaluated each implementation in a lab environment with our state learner. We learned one state machine for each combination of implementation and supported key exchange flow type, with \msgkexinit bound to that key exchange algorithm. This resulted in a total number of \nrStateMachines state machines for analysis. Detailed learning statistics can be found in the additional material in the artifacts.

For Bitvise SSH, we were forced to artificially close the connection when the learner attempted a key re-exchange, that is, when sending another \msgkexinit symbol inside an already established secure channel. This is because Bitvise SSH buffers messages sent during the key re-exchange and processes them collectively after the key re-exchange has concluded, resulting in new states with each non-KEX message sent during a key re-exchange, eventually causing a combinatorial explosion. With Tectia SSH and LANCOM LCOS, and libssh with DHGEX, our learner did not converge due to significant strict KEX violations or repeated cache conflicts. In these cases, we manually terminated the learning process after one hour and used the last hypothesis for analysis.

For every implementation, \autoref{tab:eval_ssh_impl} also includes an overview of the observed violations by category, which are described in the following subsection. We also explain every violation in greater detail in \autoref{sec:eval:sub:violations}. These results include two severe security vulnerabilities, resulting in a rogue session attack (cf.~\cite{USENIX:BauBriSch24}) against Tectia SSH and a pre-authentication remote code execution attack in Erlang SSH. We discuss these attacks in subsections~\ref{sec:eval:sub:tectia} and~\ref{sec:eval:sub:erlang}.

\newcommand{\violation}{\color{red}\faBug\xspace}
\newcommand{\noviolation}{\color{DarkGreen}\faMinus\xspace}
\begin{table*}
    \centering
    \caption{SSH server implementations and their support key exchange flow types included in the evaluation, and the observed strict key exchange violations for at least one of the flows. For all implementations, one algorithm of each key exchange algorithm flow type is learned.}
    \label{tab:eval_ssh_impl}
    \begin{tabular}{llccccccccccc}
        \toprule
        ~ & ~ & ~ & \multicolumn{5}{c}{Supported Key Exchange Flows} & \multicolumn{5}{c}{Observed Violations} \\
        \cmidrule(lr){4-8}\cmidrule(lr){9-13}
        Implementation & Version & OS & DH & DHGEX & RSA & ECDH & PQ-Hybrid & C1 & C2 & C3 & C4 \\
        \midrule
        AsyncSSH & 2.20.0 & \faLinux & \supported & \supported & \supported & \supported & \supported & \noviolation & \violation & \noviolation & \noviolation \\
        Bitvise SSH & 9.43 & \faWindows & \supported & \supported & \unsupported & \supported & \unsupported & \noviolation & \noviolation & \noviolation & \noviolation \\
        Dropbear & 2025.87 & \faLinux & \supported & \unsupported & \unsupported & \supported & \supported & \noviolation & \noviolation & \noviolation & \violation \\
        Erlang SSH & 5.2.8 & \faLinux & \supported & \supported & \unsupported & \supported & \unsupported & \noviolation & \noviolation & \violation & \noviolation \\
        Golang x/crypto/ssh & 0.37.0 & \faLinux & \supported & \unsupported & \unsupported & \supported & \unsupported & \noviolation & \noviolation & \noviolation & \noviolation \\
        LANCOM LCOS & 10.90.0126 Rel & \faNetworkWired & \supported & \supported & \unsupported & \supported & \supported & \noviolation & \noviolation & \violation & \noviolation \\
        libssh & 0.11.1 & \faLinux & \supported & \supported & \unsupported & \supported & \unsupported & \noviolation & \violation & \noviolation & \noviolation \\
        OpenSSH & 9.9p2 & \faLinux & \supported & \supported & \unsupported & \supported & \supported & \noviolation & \noviolation & \noviolation & \noviolation \\
        Tectia SSH & 6.6.5.353 & \faWindows & \supported & \supported & \unsupported & \supported & \supported & \violation & \noviolation & \violation & \noviolation \\
        TinySSH & 20250201 & \faLinux & \unsupported & \unsupported & \unsupported & \supported & \supported & \noviolation & \noviolation & \noviolation & \violation \\
        \bottomrule
    \end{tabular}
\end{table*}

\subsection{Violation Categories}

Implementing strict KEX handshake hardening measures is non-trivial, as these are not cryptographically ensured and can go unnoticed. That is because any benign peer negotiating a handshake with strict KEX enabled expects the remote peer to close the connection on optional or invalid messages, and would, therefore, shy away from sending such messages in the first place. During the evaluation, we encountered four main categories of strict KEX violations.

\begin{figure*}[t]
    \begin{subfigure}[]{.24\linewidth}
        \centering
        \frame{\includegraphics[width=\textwidth,page=2]{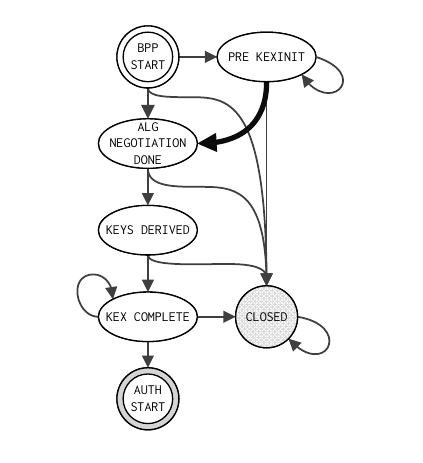}}
        \caption{C1 -- Early Message Injection}
        \label{sec:eval:fig:c1}
    \end{subfigure}
    \hfill
    \begin{subfigure}[]{.24\linewidth}
        \centering
        \frame{\includegraphics[width=\textwidth,page=3]{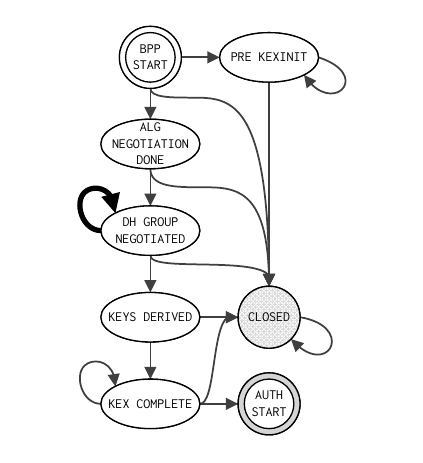}}
        \caption{C2 -- Flow-dependent Injection}
        \label{sec:eval:fig:c2}
    \end{subfigure}
    \hfill
    \begin{subfigure}[]{.24\linewidth}
        \centering
        \frame{\includegraphics[width=\textwidth,page=4]{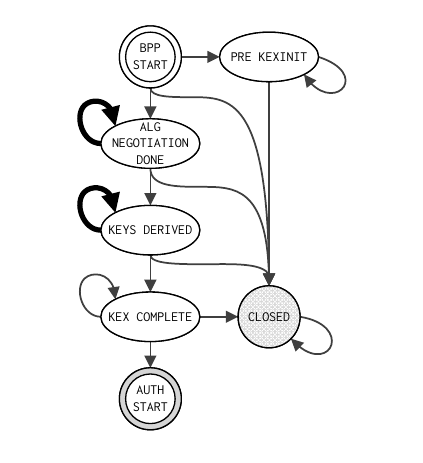}}
        \caption{C3 -- Partial Hardening}
        \label{sec:eval:fig:c3}
    \end{subfigure}
    \hfill
    \begin{subfigure}[]{.24\linewidth}
        \centering
        \frame{\includegraphics[width=\textwidth,page=5]{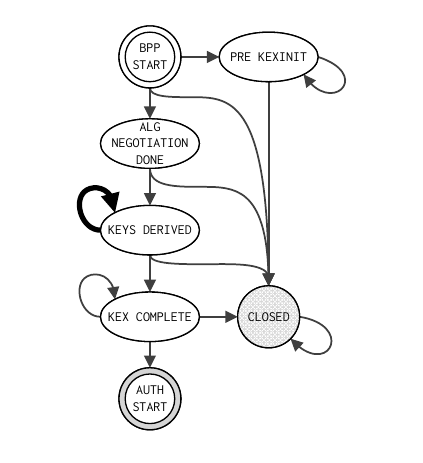}}
        \caption{C4 -- Late Message Injection}
        \label{sec:eval:fig:c4}
    \end{subfigure}
    \caption{Idealized state machines illustrating different strict KEX violation categories which were observed during the evaluation. A single implementation can be subject to multiple violation categories simultaneously. Flow-dependent injections are limited to a proper subset of all supported key exchange algorithms, while violations in other categories are independent of the negotiated key exchange algorithm.}
    \Description{The figure consists of four different state machine graphs describing different violations to the strict KEX extension. The first one, called early message injection (C1), has an additional state transition between the PRE KEXINIT and ALG NEGOTIATION DONE states. The second one has an additional self-loop inside the key exchange and is titled flow-dependent injection (C2). The third one has additional self-loops in all happy flow states and is titled partial hardening (C3). The last one has an additional self-loop just before the key exchange concludes, depicting an early restriction lifting, and is named late message injection (C4).}
\end{figure*}

\paragraph{Category 1 (C1) -- Early Message Injection (\autoref{sec:eval:fig:c1})}

As support for the strict KEX extension is signaled as part of the \msgkexinit message, enforcing strict KEX restrictions before both peers send their \msgkexinit message is impossible. Consequently, peers negotiating strict KEX for a connection must check that the remote peer's \msgkexinit was the first binary packet protocol message received in hindsight. One implementation violated this, allowing for message injection immediately after the protocol version exchange.

\paragraph{Category 2 (C2) -- Flow-Dependent Injection (\autoref{sec:eval:fig:c2})}

The sequence of allowed messages in an SSH handshake depends on the negotiated key exchange algorithm. That is, the message flow between the \msgkexinit and \msgnewkeys messages cannot be implemented with static messages, but rather requires dynamic behavior depending on the key exchange algorithms supported by the remote peer. Depending on the implementation, this can impact handshake hardening if message restrictions are only implemented for one flow type but not the other. We identified two implementations that violated this category, as they were compliant with strict KEX requirements in some but not all key exchange flow types.

\paragraph{Category 3 (C3) -- Partial Hardening (\autoref{sec:eval:fig:c3})}

When implementing handshake hardening measures as a blocklist rather than an allowlist, for example, by explicitly filtering \msgignore and \msgdebug, implementors may overlook optional message types or not consider unknown message IDs. Additionally, blocklists must be changed dynamically to prevent expected message types from being accepted out of order. We found three implementations that missed at least one unexpected or invalid message when enforcing strict KEX.

\paragraph{Category 4 (C4) -- Late Message Injection (\autoref{sec:eval:fig:c4})}

Restrictions by strict KEX must be enforced until both parties have concluded the key exchange by sending \msgnewkeys. Although no strict order for \msgnewkeys messages is enforced in the specification, for all key exchange flow types except RSA, the server is the first party capable of computing the session keys and, thus, sending the \msgnewkeys message. Therefore, servers must take care not to lift strict KEX restrictions after sending \msgnewkeys, but after receiving \msgnewkeys from the client. We found two server implementations that allowed otherwise forbidden messages before the key exchange concluded.

\subsection{Strict Key Exchange Violations}
\label{sec:eval:sub:violations}

During our evaluation, we identified a total of \nrViolatingImplementations server implementations that failed to meet the requirements set by strict KEX. Golang's x/crypto/ssh library and OpenSSH are the only implementations we analyzed that implement a strict KEX compliant with the specification and immediately close the connection.

\begin{figure}
    \centering
    \frame{\includegraphics[width=.7\columnwidth,page=6]{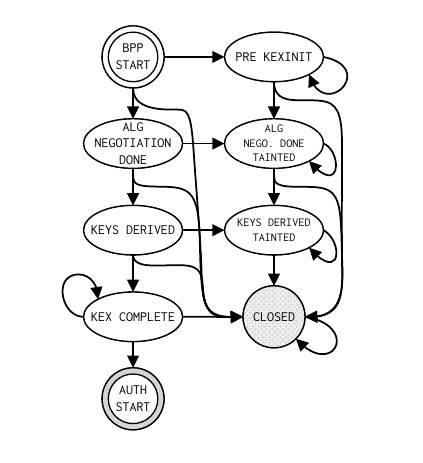}}
    \caption{Taint-based implementation of strict KEX used by libssh and Bitvise SSH. When an unexpected or invalid message is received, the internal state transitions into a tainted state, which eventually causes the connection to terminate. We do not consider taint-based implementations an inherent violation of strict KEX requirements if implemented correctly.}
    \Description{A state machine depicting a taint-based implementation of strict KEX used by libssh and Bitvise SSH. The protocol flow starts in the BPP START state and then continues down the happy flow of algorithm negotiation, key exchange and derivation, and activation of the secure channel. When an unexpected or invalid message is received, the state transitions into a tainted mirror state that eventually causes the connection to terminate.}
    \label{sec:eval:fig:taint}
\end{figure}

Bitvise SSH and libssh implement strict KEX with a different approach by maintaining an internal taint flag that is set whenever an unexpected or invalid message is received during key exchange. Once the key exchange concludes, the server terminates the connection if the taint flag is set and strict KEX is active, resulting in a state machine similar to the one shown in \autoref{sec:eval:fig:taint}. While not strictly compliant with the specification, this approach prevents any handshake violating the strict KEX requirements from continuing into the secure channel. Therefore, we do not consider taint-based implementations an inherent violation of the requirements of strict KEX if implemented correctly without additional violations.

\paragraph{Proof-of-Concept} 

For each implementation with strict KEX violations, we chose one violating message trace and verified it by executing it against the server, outside the state learning routine, with our SSH implementation. A violation is verified if the client can successfully establish a secure channel and reach authentication. We provide all proof-of-concepts as part of our artifacts.

\paragraph{AsyncSSH} 

For AsyncSSH, we found a minor violation of category C2 in the Diffie-Hellman group exchange flow type. In particular, AsyncSSH does not limit the number of \msgkexdhgexrequest messages but instead responds to each group request with a finite field to use within the Diffie-Hellman key exchange. This behavior is not a security vulnerability in AsyncSSH because the server correctly updates both the DH group for the key exchange and the one included in the signed exchange hash simultaneously.
    
\paragraph{Dropbear} 

Dropbear violates strict KEX by lifting key exchange restrictions too early, resulting in a category C4 violation. This allows for message injection after the server signals \texttt{NEWKEYS}, but before the client does so as well. At this point, the connection is partially encrypted; that is, messages sent by the client are still unencrypted, while messages sent by the server are encrypted using the derived algorithms and keys. This causes messages that prompt a server response to increment the SQN send counter after the sequence number is reset by the strict key exchange extension, ultimately allowing an attacker to offset sequence numbers in the secure part of the protocol from within the unencrypted one. This directly violates the strict key exchange's goal to separate unencrypted and encrypted sequence numbers. Message injection is limited to a subset of messages, specifically any message ID less than 50---the first message ID of the authentication protocol.

\paragraph{Erlang SSH} 

Erlang SSH shows violations of category C3 by accepting connection protocol messages during the handshake. Additionally, optional messages such as \msgignore and \msgdebug do not cause the connection to terminate. A closer inspection of the violations revealed a remote code execution vulnerability, which we describe in greater detail in \autoref{sec:eval:sub:erlang}.

\paragraph{LANCOM LCOS} 

Our findings indicate significant violations of category C3 in LANCOM LCOS. In fact, the implementation replied to all unknown message IDs and message IDs greater than or equal to $50$ with \msgunimplemented messages. Interestingly, LANCOM LCOS does not appear to support \msgdebug messages and considers them unknown, as evidenced by the \msgunimplemented messages sent in response. However, other optional messages, such as \msgignore, are correctly handled and cause the connection to terminate.

\paragraph{libssh} 

Like Bitvise SSH, libssh uses the taint-based approach depicted in \autoref{sec:eval:fig:taint} to implement the strict KEX extension. The taint flag is only set for messages of the transport layer; messages with a message ID greater than or equal to $50$ cause the connection to terminate immediately during the handshake. However, libssh exhibits a violation of category C2 by failing to enforce strict KEX requirements for DHGEX flow types.

\paragraph{Tectia} 

Tectia SSH does not correctly handle optional messages such as \msgignore and \msgdebug, causing a violation of category C3. Additionally, Tectia SSH allows these messages to be sent before the \msgkexinit without causing the connection to terminate as soon as the client's \msgkexinit message is received. Therefore, this violation extends to category C1 as well. We also identified severe bugs concerning the handling of user authentication requests, allowing for a rogue session attack. We describe this attack in \autoref{sec:eval:sub:tectia}.

\paragraph{TinySSH} 

TinySSH lifts strict key exchange restrictions too early, and, therefore, exhibits a violation of category C4, allowing for message injection at the same stage as Dropbear. However, message injection, in this case, is not limited to a subset of transport layer protocol messages. Instead, all messages, except \texttt{DEBUG}, \texttt{IGNORE}, and \texttt{DISCONNECT}, can be inserted without the server terminating the connection. Furthermore, TinySSH does not send a response to any of these messages, maintaining sequence number separation between the encrypted and unencrypted parts of the protocol.

\subsection{Rogue Session Attack against Tectia SSH}
\label{sec:eval:sub:tectia}

\begin{figure}
    \centering
    \frame{\includegraphics[width=\linewidth]{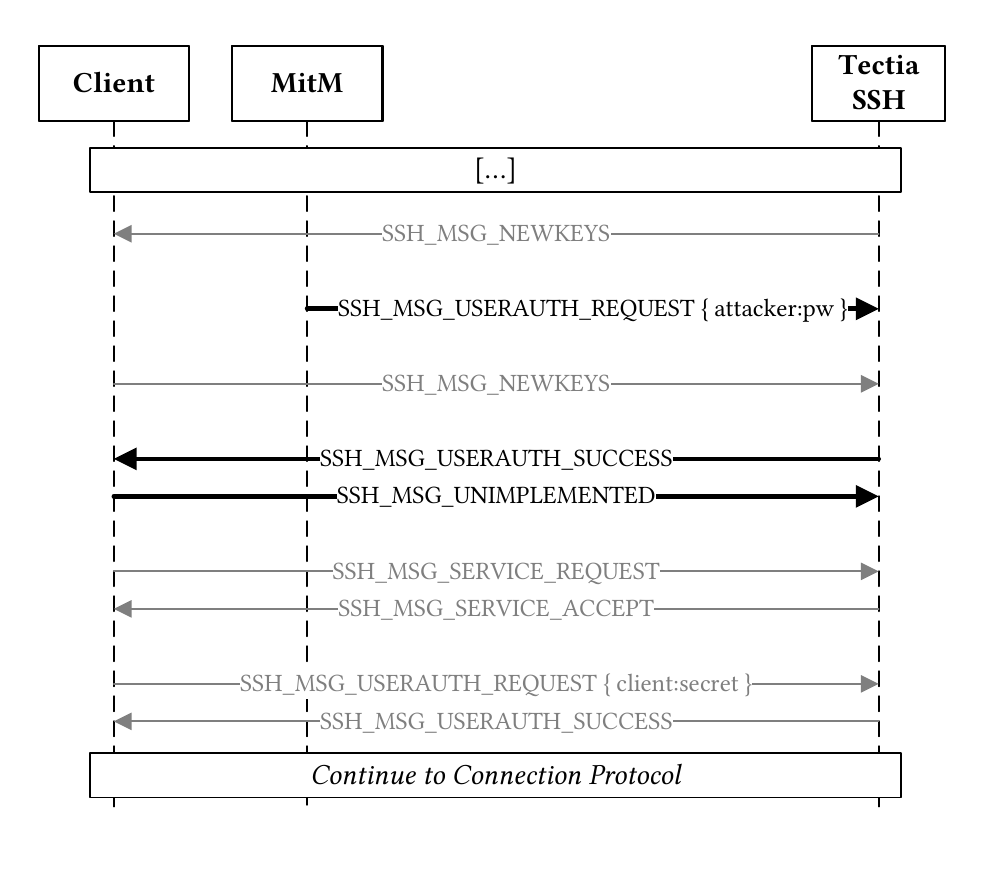}}
    \caption{Rogue session attack targeting clients connecting to a Tectia SSH server with strict KEX active. A MitM attacker injects a rogue authentication request just before the handshake concludes, which is possible due to the sequence number reset. The client must be lenient and not terminate the connection when receiving an early authentication success. The second authentication request causes a repeated authentication success without changing the authenticated user.}
    \Description{The figure depicts a protocol flow between a client and a Tectia SSH server, with a MitM attacker in between. The initial messages of the SSH protocol are left out, and the flow starts with the NEWKEYS message from the server to the client. Before the client sends its NEWKEYS, the MitM attacker injects a rogue authentication request containing the attacker's credentials. The server responds to this request with a success message, which the client then responds to as \msgunimplemented, as the message is not valid at the moment. The connection then continues normally.}
    \label{sec:eval:fig:tectia}
\end{figure}

Tectia SSH is the official SSH implementation of SSH.COM, a company founded by Tatu Ylönen, the inventor of the SSH protocol. According to the company's website, ``Tectia SSH is trusted by 40\% of Fortune 500 companies and 4/5 of the world's largest banks.''\footnote{\url{https://info.ssh.com/tectia-server-client-datasheet-download}, accessed: 2025-04-14}

Upon closer examination of the extracted state machines, we find that Tectia SSH allows \msguserauthrequest messages to be sent during the handshake. All \msguserauthrequest messages are seemingly buffered until after the handshake concludes, where the server responds to these authentication requests. Tectia SSH also allows immediate user authentication without the client switching protocols using a service request first. After successful user authentication, Tectia SSH continues to answer service and user authentication requests without terminating the connection.

All of these observations can be combined into the rogue session attack depicted in \autoref{sec:eval:fig:tectia}, which is similar to the one described in~\cite[Section~6.2]{USENIX:BauBriSch24}. However, the attack becomes significantly less complicated due to the sequence number reset of strict KEX and the fact that Tectia SSH handles service and authentication requests inside the connection protocol. To exploit this vulnerability, a MitM attacker can inject a user authentication request containing the attacker's credentials before the server receives the client's \msgnewkeys.

This authentication request will be processed immediately by the server, and the client will likely ignore the \msguserauthsuccess message sent as a response, as strict KEX restrictions have already been lifted. We confirmed that this is the case for OpenSSH and PuTTY, two common SSH clients. The client then sends \msgservicerequest and \msguserauthrequest messages, the latter most likely of type \texttt{none}, to which the server responds positively as the session is already authenticated. In fact, this behavior violates~\cite[Section 5.1]{rfc4252}, which mandates that the \msguserauthsuccess message must not be sent more than once. This violation causes both peers to resynchronize their state at the beginning of the connection protocol, while the authenticated account is the attacker's. This finding has been assigned \tectiacve with an estimated CVSSv3 score of 6.8.

We implement a proof-of-concept based on the Terrapin artifacts,\footnote{\url{https://github.com/RUB-NDS/Terrapin-Artifacts}, accessed: 2025-04-14} and confirmed that we can successfully log a victim using a recent OpenSSH client into a shell account under the attacker's control on a Tectia SSH server, enabling phishing attacks.

\subsection{Remote Code Execution in Erlang SSH}
\label{sec:eval:sub:erlang}

\begin{figure}
    \centering
    \frame{\includegraphics[width=\linewidth]{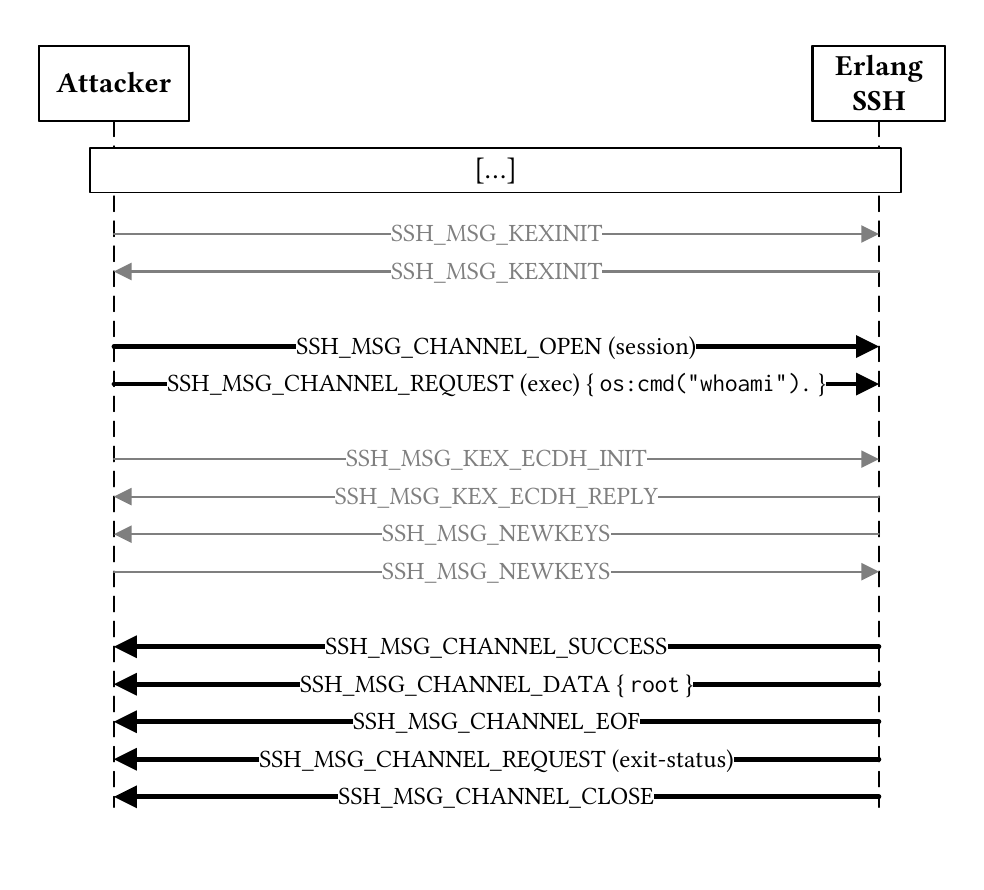}}
    \caption{Remote code execution in Erlang SSH. A malicious actor can open a logical channel from the connection protocol early and send an \texttt{exec} request. The server will respond to this request at the beginning of the secure channel with a success message, the command's output and exit status, and subsequently close the channel.}
    \Description{The figure depicts a remote code execution in the server implementation of Erlang SSH. The attacker opens a channel and executes a command (here: os:cmd("whoami")) immediately after the KEXINIT messages. The key exchange continues normally, after which the server sends the response to the attacker's command (here: root) as well as various channel-related messages.}
    \label{sec:eval:fig:erlang}
\end{figure}

Erlang SSH is part of the Erlang Open Telecom Platform, which is middleware for implementing highly reliable, distributed systems, such as telecommunication and electronic payment systems. Although these systems are usually not directly accessible on the Internet, Shodan.io listed 169 servers in April 2025 that return an \texttt{SSH-2.0-Erlang} banner.

In the learned state machine, we found that Erlang SSH violates strict KEX restrictions by allowing the client to send \msgchannelopen messages immediately after the \msgkexinit messages, which is a violation of category C3. However, instead of a self-loop, the extracted state machine indicates that sending \msgchannelopen triggers a state change that can be reverted by sending a \msgchannelclose message, raising suspicions whether messages are just accepted or actually processed. Furthermore, after sending \msgchannelopen, the server seemingly accepts \msgchannelrequest messages without closing the connection. However, the state learner has not received an immediate response to any of these messages.

Upon further investigation of this behavior using our SSH implementation, we observed several channel-related server responses after the handshake but before client authentication, corresponding to the channel-related client requests sent during the handshake. As shown in \autoref{sec:eval:fig:erlang}, sending a \msgchannelrequest of type \texttt{exec} during the handshake results in a \msgchannelsuccess message alongside a \msgchanneldata message containing the output of the Erlang code included in the request. As all of this happens before client authentication, this behavior resembles a remote code execution vulnerability that can be exploited without privileges or user interaction. Our finding has been assigned \erlangcve with an estimated CVSSv3 score of 10.0.

We describe a proof-of-concept protocol flow in an XML file that can be used with our SSH implementation. We confirmed that our proof-of-concept allows the execution of a shell command without prior authentication in the context of the Erlang SSH daemon. The proof-of-concept is included as part of our artifacts.

\section{Related Work}

\paragraph{State Learning}

Applying model learning to analyze the security of cryptographic protocols was first proposed by de Ruiter~et~al.~\cite{USENIX:dRuPol15} in the context of the Transport Layer Security (TLS) protocol. The authors coined the term \emph{protocol state fuzzing} for this approach, as protocol state transitions are randomly learned in black box testing, resulting in a state machine that can be analyzed manually or with model checking. Various studies further extended the analysis for TLS and Datagram TLS~\cite{SecITSys:dRu16,yadav2018tls,USENIX:FJMRSS20,CCS:STVHBC22,ESORICS:RasLevDeb22}. Karim~et~al. studied the Bluetooth Low Energy (BLE) protocol~\cite{SP:KIHB23} and proposed a methodology to learn phases of the protocol separately but merge them into one comprehensive state machine later. In 2024, Tu~et~al.~\cite{USENIX:TIRDWW24} studied the state machines of 5G basebands using a novel approach combining passive and active state learning and by leveraging the models of different basebands for their equivalence oracle. Dong~et~al. introduced the concept of guiding state machines from protocol specifications to implement a mutation-based testing strategy for 4G LTE and BLE devices in~\cite{CCS:RWTITD24}. They applied a related concept to 5G core network implementations in~\cite{dong2025corecrisis} using learned state machines, thereby deriving test cases that can evaluate protocol states in greater detail.

In 2017, Fiter\u{a}u-Bro\c{s}tean~et~al. first applied state learning to SSH~\cite{SPIN:fiterauSshModelChecking}. They extracted state machines of three implementations and applied model-checking techniques to test if each state machine achieves certain functional and security properties. Their learner commenced after the exchange of the ASCII banner; however, they further partitioned the protocol into three layers (transport layer, authentication layer, and connection layer) and learned each layer individually with a distinct alphabet designed for the respective protocol phase. In 2023, Fiter\u{a}u-Bro\c{s}tean~et~al. further expanded their approach~\cite{NDSS:FJST23} and proposed a methodology for the automated validation of state machines through manually modeled bug patterns. For the SSH part of this paper, they used the three extracted state machines from~\cite{SPIN:fiterauSshModelChecking} as artifacts.

\paragraph{SSH Security Analyses} 

Various studies have presented practical analyses on the security of the SSH protocol, its implementations, and deployments on the Internet. For instance, Albrecht~et~al.~\cite{SP:AlbPatWat09} demonstrated how an attacker can use encrypted length fields to uncover the plaintext of a message. Song~et~al.~\cite{USENIX:SonWagTia01} showed that keystroke timing attacks allow an attacker to deduce passwords typed by a user during interactive authentication in a shell. Bhargavan and Leurent~\cite{NDSS:BhaLeu16} analyzed the impact of SHA-1 collisions on the session transcript of SSH. In a study evaluating the host keys of SSH deployments on the Internet, Heninger~et~al.~\cite{USENIX:HDWH12} found that implementation errors and lack of entropy in generated key material leave some hosts vulnerable to impersonation attacks. Bäumer~et~al.~\cite{USENIX:BauBriSch24} presented the Terrapin attack, a prefix truncation attack against some SSH cipher modes. They also give a rogue session attack similar to the one in \autoref{sec:eval:sub:tectia}. The countermeasure they propose is the subject of our study. Formal proofs for SSH have been contributed by Bellare~et~al.~\cite{CCS:BelKohNam02}, Bergsma~et~al.~\cite{CCS:BDKSS14}, Paterson and Watson~\cite{EC:PatWat10}, Williams~\cite{IMA:Williams11}, and Hansen~\cite{hansen2020cryptographic}.
\section{Discussion}

\subsection{Mitigations}
\label{sub:mitigations}

First and foremost, the vulnerability in Erlang SSH can be mitigated by disallowing all messages unrelated to the handshake phase before the completion of the first key exchange. This vulnerability is a straightforward layer violation, and thus easy to mitigate by safeguarding the implementation against unexpected message IDs.

\paragraph{Avoiding Strict KEX Violations}

Our analysis, as well as the subsequent responsible disclosure process, indicates that there is no single solution for preventing strict KEX violations due to fundamental differences in the code base of different implementations and the multitude of potential pitfalls. Instead, thorough testing of an implementation is required to rule out the presence of such violations. We observed that most implementers encountered significant challenges in verifying their patches because the implementation could not establish connections that violated strict KEX for negative testing. To resolve this issue, implementers may use our artifacts to verify the correctness of their strict KEX implementations.

\paragraph{Full Transcript Verification}

We observe that violations of the handshake hardening measures of strict KEX go unnoticed because they are not cryptographically enforced. Incorporating a full transcript verification, similar to the \texttt{verify\_data} approach of TLS 1.3~\cite{rfc8446}, would be one direction to harden the SSH handshake further against attacks like Terrapin. Bäumer~et~al. already discussed the applicability of a full transcript hash to mitigate the Terrapin vulnerability~\cite{USENIX:BauBriSch24}. Full transcript verification is an alternative approach to mitigating the Terrapin vulnerability, intended to supersede strict KEX, and does not prevent strict KEX violations. Our results emphasize the benefits of a full transcript hash; however, we also acknowledge the challenges involved in applying this idea.

Implementing a full transcript verification first requires a canonicalized message order since, in SSH, some messages do not have a well-defined order and may be sent simultaneously. For example, some servers may wait until they receive the client's \msgkexinit message, while others send their \msgkexinit immediately after the banner exchange. Similar to strict KEX, the client and server could then negotiate the use of full transcript verification in the \msgkexinit messages as pseudo-algorithm identifiers. The use of an extension as part of the \msgextinfo message is not possible, as this message is only exchanged after the handshake is concluded. If the extension is negotiated, the client extends its service request to include the value $MAC(K_x, H_c || H_s)$ where $K_x$ is a separate MAC key derived during key derivation with a different label for the sole purpose of authenticating the handshake. $H_c$ and $H_s$ are the unidirectional transcript hashes sent by the client and server, respectively, beginning at the very start of the connection up to and including the parties' \msgnewkeys message and computed with the hash function of the negotiated key exchange. Upon receiving the client's \msgservicerequest containing the transcript MAC, the server verifies it and aborts the connection if verification fails. Similarly, but using a different MAC key $K_y$, the server authenticates the transcript to the client in its \msgserviceaccept message, ensuring that both client and server acknowledged the transcript.

\paragraph{Limitations of Transcript Verification}

Note that the transcript verification approach does not ensure the correct rejection of invalid messages. For example, consider the Erlang vulnerability described in \autoref{sec:eval:sub:erlang}, which is triggered by an attacker sending an invalid channel open and channel request message. In this scenario, the attacker assumes the role of a client, giving them control over any session transcripts they issue. Neither including nor omitting these messages from the transcript on the server side would prevent the attack, as the attacker would always be able to produce a matching transcript of their own.

\subsection{Performance}
\label{sec:eval:sub:equiv}

\begin{table}
    \centering
    \small
    \caption{Comparison of different equivalence algorithms when learning OpenSSH 9.9p2 with strict KEX enabled and an alphabet size of $|\mathcal{A}| = 86$.
    For random oracles, we limited the number of queries to $1{,}000{,}000$. For the Wp-method, we chose a lookahead of $2$ and expected $10$ states overall. For our happy flow mutation, we allowed up to $n = 2$ message insertions. Queries answered by the cache are excluded. A dagger symbol (\textdagger) indicates that the equivalence oracle did not return after a reasonable time and was terminated manually.}
    \label{sec:discussion:sub:perf:tab:comparison}
    \begin{tabular}{lrccrr}
        \toprule
        ~ & ~ & \multicolumn{2}{c}{\# States} & \multicolumn{2}{c}{\# Queries} \\ \cmidrule{3-4}\cmidrule{5-6}
         Algorithm & Time [s] & Total & Key Ex. & Total & Final  \\ \midrule
         Random words & $3{,}189$\phantom{\textsuperscript{\textdagger}} & 8 & 6 & $45{,}466$ & $21{,}163$ \\
         Random Wp & $37{,}367$\phantom{\textsuperscript{\textdagger}} & 38 & 6 & $792{,}680$ & $745{,}542$ \\
         Wp-method & $128{,}832$\textsuperscript{\textdagger} & 18 & 6 & $4{,}533{,}240$ & $369{,}529$ \\
         \midrule
         This work & 206\phantom{\textsuperscript{\textdagger}} & 9 & 6 & 8,906 & 3,769 \\ \bottomrule
    \end{tabular}
\end{table}

Our methodology is specifically tailored to the detection of strict KEX violations. To this end, our proposed equivalence oracle focuses on in-depth verification of handshake states but applies less scrutiny to post-handshake states, which are not in the scope of our study. As shown by our results, this approach allows us to detect issues in real-world implementations of the strict KEX mechanisms; however, it also allows us to extract state machines quickly. To illustrate this, we compared our equivalence oracle to other strategies used in prior work applying state learning to network protocols. Specifically, we use the implementation provided by LearnLib. Our results in \autoref{sec:discussion:sub:perf:tab:comparison} indicate that our equivalence oracle requires significantly fewer queries, and thus execution time, than generic approaches, while still discovering all relevant key exchange states. In contrast, generic approaches typically require more queries to discover additional authentication protocol states.

\subsection{Message Injection in SSH}

As seen in \autoref{sec:eval:sub:tectia}, incomplete implementations of strict KEX can ease message injection into the handshake, potentially exposing client and server implementation flaws to a MitM attacker. This issue becomes visible when comparing the rogue session attack against Tectia SSH with the original rogue session attack against AsyncSSH by Bäumer~et~al.~\cite{USENIX:BauBriSch24}. Without strict KEX, Bäumer~et~al. resort to the Terrapin attack's prefix truncation to enable a single message injection into the handshake. With strict KEX enabled, this is no longer necessary due to the sequence number reset. Therefore, if handshake hardening measures are faulty, an attacker can inject an arbitrary number of messages into the handshake.

However, message injection without strict KEX does not necessarily involve performing a successful prefix truncation attack. A straightforward yet effective approach to inserting messages into the handshake is to replace optional unencrypted messages from the client or server with meaningful ones. Because messages are replaced rather than inserted, this does not cause sequence numbers to become misaligned, making this injection method independent of the algorithm choice.

Another injection vector can be found in the specification of the AES-GCM encryption algorithms for SSH~\cite{rfc5647}. This specific AES-GCM construction does not rely on sequence numbers but instead uses an 8-byte invocation counter as part of the nonce to prevent packet deletion, reordering, or injection. The counter is initialized using output from the key derivation. While this is a direct violation of~\cite{rfc4253}, which requires the sequence number to be used in the computation of the integrity value, the specification's authors claim that the sequence numbers are equal to the invocation counter minus a constant offset. However, sequence numbers are never used to derive the initial value of the invocation counter, allowing for trivial message injection into the initial SSH handshake.

To the best of our knowledge, message injection for AES-GCM, in particular, and by replacement of optional messages, has not been previously described elsewhere and is also not included in the security considerations of the specification. As these vectors apply to SSH in general, some of our findings may apply to implementations without strict KEX, for example, in the case of the rogue session attack against Tectia SSH. We therefore appeal to the IETF sshm working group to include the specification of a full handshake verification in their charter.

\subsection{Limitations}
\label{sec:discussion:sub:limitations}

\paragraph{Mapper Assumptions}

Most protocol messages are designed to affect the protocol state, for example, by advancing the handshake. Especially when sending such messages outside the protocol's expected order, it is unclear if the recipient actually processes it in whole or discards it, leaving the internal state unaffected. Hence, the implementers must define assumptions for the mapper on how to proceed when executing a symbol out of order. For example, our mapper assumes all \texttt{NEWKEYS} messages to set a pending cipher state as the new active state. If no cipher state has been negotiated yet because the learner queries the symbol too early, we assume a \texttt{none} cipher state using no message protection. We further assume each \texttt{NEWKEYS} will reset the sequence numbers as expected by strict KEX. However, the recipient may not always change its cipher state or reset its sequence numbers. In this case, the session state of our mapper and the SUL would become misaligned, and the following encrypted message sent would get rejected as invalid simply because we changed our state while the peer did not. Incorrect anticipation of the peer's behavior can introduce learning artifacts in the obtained state machines, manifesting as additional states. While these states are technically valid, they hinder comparability between different mappers unless the underlying assumptions are identical. This may also cause an implementation to behave non-deterministically, for example, when message decryption fails and returns unpredictable results instead.

\paragraph{Non-Determinism}
\label{par:nondeterminism}

While most of our queries are deterministic due to the use of a reliable transport protocol, some queries exhibit non-deterministic behavior. Non-determinism poses a significant challenge to our cache, resulting in cache conflicts that necessitate restarting the learning algorithm. Continuing without a restart may cause the algorithm to produce inaccurate results or to terminate unexpectedly due to contradictory query outputs. However, some perceived non-determinism may actually stem from race conditions due to timeouts defined in the implementations. For example, the \texttt{LoginGraceTime} configuration option of OpenSSH causes the server to close an unauthenticated connection once an internal timer expires. As our state learner does not operate in constant time, we may query varying numbers of symbols in a given time frame. Consequently, we would sometimes observe the closed connection sooner (in terms of queried symbols) and sometimes later, making cache conflicts very likely. Ultimately, the number of cache conflicts renders our approach infeasible, as the learning process never converges. While earlier works used majority voting (best out of three) for all queries~\cite{CCS:HKICB21,NSDI:MHYBBSWYZ19}, we only fall back to this strategy to resolve cache conflicts, but then use a larger voting population of 13 to handle queries with a high degree of non-determinism.

Another source of seemingly non-deterministic behavior is the probabilistic nature of cryptographic protocols involving nonces and randomness in general. In that case, we may not be able to reproduce the same behavior when executing a chain of symbols twice, as the individual abstract symbols only cover semantic equality, not byte-wise equality.

\section{Conclusion and Future Work}

Throughout this paper, we learned a total of \nrStateMachines state machines of \nrImplementations server implementations. We find that \nrViolatingImplementations out of \nrImplementations server implementations suffer from at least one strict KEX violation, allowing us to answer our first research question in the negative. We further show that for two implementations, we can exploit violations to obtain severe security vulnerabilities, such as a rogue session attack against Tectia SSH and remote code execution in Erlang SSH, answering our second research question.

We grouped the violations into four different categories, each corresponding to a potential pitfall when implementing the handshake hardening measures of strict KEX. We also observed two implementations that use a taint-based implementation of strict KEX, which, while not strictly adhering to the specification, can achieve the same security level if implemented correctly.

Our findings should act as a reminder to protocol designers that defenses should be cryptographically enforced, if possible. Non-cryptographic countermeasures can be tricky to implement, and violations of the specification may not be apparent and require extensive testing to be found. This is especially true for two-part countermeasures consisting of cryptographic and non-cryptographic parts, where failing to implement one part can void the security benefits of the other or even allow for new attack vectors.

\paragraph{Future Work}

Our work strictly analyzes the state machines of SSH server implementations. In the future, examining the state machine of SSH client implementations with and without strict KEX may reveal new, previously unknown weaknesses and vulnerabilities. This is especially true because the SSH specification does not prevent servers from trying to open channels or spawning shells. To do this, future work may resort to techniques like snapshotting to reduce the learning times of deeper protocol levels.

Future work may also improve on the limitations of this work. For mapper assumptions, these may be derived by analyzing implementations or performing an automated test suite before state learning. By doing so, state learning may become applicable to a broader range of SSH implementations, both with and without strict KEX enabled. Identifying possible sources of non-determinism in SSH implementations can be valuable in eliminating or at least reducing cache conflicts, resulting in more accurate state machines.

\begin{acks}
Fabian Bäumer and Marcel Maheren were supported by the German Federal Ministry of Research, Technology and Space (BMFTR) project ``Combinatorial testing of TLS libraries at all levels (KoTeBi)'' (16KIS1557). Marcus Brinkmann was supported by the Deutsche Forschungsgemeinschaft (DFG, German Research Foundation) under Germany's Excellence Strategy - EXC 2092 CASA - 390781972.

\end{acks}

\bibliographystyle{ACM-Reference-Format}
\balance
\bibliography{bib/abbrev2,bib/crypto,bib/rfc,bib/paper}

\end{document}